\documentclass[aps,prl,twocolumn,groupedaddress]{revtex4}
\usepackage{graphicx}
\usepackage{amsmath}
\usepackage{amsfonts}
\usepackage{amssymb}
\usepackage{epstopdf}

\graphicspath{{./Figures/}} %Gives LaTeX additional folders to look for image files.
\begin{document}

%\preprint{}

\title{Myocardial Architecture and Patient Variability in Clinical Patterns of Atrial Fibrillation}


\author{Kishan A. Manani$^{1,2,3}$, Kim Christensen$^{1,3}$, Nicholas S. Peters$^{2}$}
%\thanks{}
\affiliation{$^{1}$The Blackett Laboratory, Imperial College London, London SW7 2BW, United Kingdom \\ 
$^{2}$National Heart and Lung Institute, Imperial College London, London W12 0NN, United Kingdom \\
$^{3}$Centre for Complexity Science, Imperial College London, London SW7 2AZ, United Kingdom}
\newcommand{\PCoupling}{\nu}
\newcommand{\PDefect}{\delta}
\newcommand{\Pcoupling}{\nu}
\newcommand{\Pdefect}{\delta}
\newcommand{\Dysf}{\epsilon}
\newcommand{\RP}{\tau}
\newcommand{\lm}{\ell_{\text{min}}}
\newcommand{\pv}{p_{\nu}}
\newcommand{\dxprime}{\Delta x^{\prime}} 
\newcommand{\dtprime}{\Delta t^{\prime}}
\newcommand{\dyprime}{\Delta y^{\prime}}
\newcommand{\RPprime}{\tau^{\prime}}
\newcommand{\CVxprime}{{\theta_{x}}^{\prime}}

\date{\today}

\begin{abstract}
Atrial fibrillation (AF) increases the risk of stroke by a factor of four to five and is the most common abnormal heart rhythm. The progression of AF with age, from short self-terminating episodes to persistence, varies between individuals and is poorly understood. An inability to understand and predict variation in AF progression has resulted in less patient-specific therapy. Likewise, it has been a challenge to relate the microstructural features of heart muscle tissue (myocardial architecture) with the emergent temporal clinical patterns of AF. We use a simple model of activation wavefront propagation on an anisotropic structure, mimicking heart muscle tissue, to show how variation in AF behaviour arises naturally from microstructural differences between individuals. We show that the stochastic nature of progressive transversal uncoupling of muscle strands (e.g., due to fibrosis or gap junctional remodelling), as occurs with age, results in variability in AF episode onset time, frequency, duration, burden and progression between individuals. This is consistent with clinical observations. The uncoupling of muscle strands can cause critical architectural patterns in the myocardium. These critical patterns anchor micro-re-entrant wavefronts and thereby trigger AF.  It is the number of local critical patterns of uncoupling as opposed to global uncoupling that determines AF progression. This insight may eventually lead to patient specific therapy when it becomes possible to observe the cellular structure of a patient's heart. 
\end{abstract}

%\keywords{}

\maketitle


%\section{Introduction}

A key challenge in the mathematical modelling of diseases is to link microscopic variation in individuals (e.g., genetic, metabolic or tissue structure) to variation in disease outcomes (e.g.,  the occurrence, recurrence or persistence of AF).  In this paper we show how variation in microstructure affects the behaviour of AF, suggesting a single mechanism for the origin of clinically observed variability of AF behaviour. 
Atrial fibrillation is characterised by the apparently random propagation of multiple activation wavefronts in atrial muscle (myocardium). This gives rise to AF episodes of variable duration. Typically, short self-terminating episodes become longer with time until they do not terminate spontaneously. Current clinical guidelines (ACA/AGA/ESC) define AF by the episode duration as paroxysmal ($< 7$ days), persistent ($> 7$ days), long-standing persistent  ($> 1$ year) and permanent (clinical decision to not treat) \cite{camm2010guidelines}. However, AF episodes will in fact lie on a continuum of durations. The natural history of AF is usually discussed using this classification scheme which by its technical definition allows for progression of paroxysmal to persistent but not the reverse. However, this classification scheme becomes problematic in cases where episodes lasting longer than 7 days terminate and are followed by episodes shorter than 7 days which is observed to occur frequently in patients \cite{Veasey2015}. Indeed, Sugihara \textit{et al} \cite{Sugihara2015}. could not consistently apply these guidelines to their continuously monitored patients because AF episode durations were observed to remit from more than 7 days to less than 7 days. Hence, they defined a different classification scheme based on AF burden (fraction of time in AF) to describe their observations rather an arbitrary 7 day cut-off to distinguish between patients.  It has been suggested that AF induces atrial electrophysiological changes (e.g., action potential duration shortening) and microstructural changes (e.g., fibrosis or gap junctional uncoupling), which promote further AF. This self-perpetuation has been termed “AF begets AF” \cite{Wijffels1995}. Whilst fibrosis promotes AF, the quantitative relationship and the mechanism by which fibrosis promotes AF is not fully understood \cite{DeJong2011}.  

Sugihara \textit{et al.} \cite{Sugihara2015} monitored AF patients continuously in a long term study (1031 cumulative patient-years, mean 3.2 years per patient) using dual chamber permanent pacemakers. It was observed that progression to persistent AF was not inevitable, that is,  some patients remained paroxysmal for the duration of the long term follow-up and some patients' AF burden (fraction of time in AF) could remit from 100 \% to less than 100 \% and relapse to 100\% again. Indeed it has been observed that some patients do not progress from paroxysmal to persistent AF, using current clinical guidelines, after as many as 22 years \cite{Kottkamp2013}. Veasey \textit{et al.} \cite{Veasey2015} also used continuous monitoring data to show that after a mean 7 year follow up, 35 \% of patients that were initially classified as persistent AF using the current clinical guidelines were re-classified as paroxysmal AF. Other research has shown that the time course of AF is seen to vary between patients with similar fibrosis burden: some patients progress rapidly from paroxysmal AF to persistent AF (on the order of months) whilst other patients do not progress at all (measured over decades) \cite{Kottkamp2013}. Furthermore, patients with a high fibrosis burden can remain paroxysmal and those with low fibrosis burden can be in persistent AF \cite{Kottkamp2013,Oakes2009,Teh2012,Boldt2004}.

 Thus we have the following recent clinical observations: (1) AF burden does not inexorably increase and can even spontaneously decrease therefore not all patients appear to progress to persistent AF, (2) persistent AF can remit to paroxysmal AF, (3) the common conception that fibrosis correlates with AF progression needs to be reconciled with the observed variability in AF burden for patients with similar levels of fibrosis. Individually and collectively these studies challenge the contemporary view of how AF evolves. Although it has been suggested that different pathological processes (mitral valve disease, diabetes etc.) occurring in different patients may contribute to variability in AF progression \cite{Schotten2011}, it is fair to say that there is no understanding of what causes the clinical observations summarised above.

The clinical patterns of AF are studied in the domain of populations on long time-scales (months to years), whereas the microstructure of myocardium is often studied in “wet labs” within the domain of cellular electrophysiology on short time-scales (form seconds to hours). These two vastly different time scales cannot be related experimentally. Similarly, many models of AF are computationally intensive due to their complexity. As a result only short time periods (seconds or minutes) have been investigated. Thus, these models cannot address questions pertaining to the long time scales of disease progression. Previous work by Chang \textit{et al.} \cite{Chang2015} has explored the two time scales by modelling AF as a simple binary process which flips between normal sinus rhythm (SR) and arrhythmia at “patient” specific rates. But, this study does not address the question of the microscopic origin of variability in clinical observations.

We use a very simple computational model to link the two time domains. The model is specifically designed to address the hypothesis that the stochastic nature of transversal uncoupling is an important factor in the temporal patterns of AF. We model a “patient” by simulating “patient” specific tissue using a simple stochastic process, and then assess the resulting temporal AF patterns. 

 The incidence of AF increases with age and is strongly associated with the accumulation of fibrosis \cite{DeJong2011}. In this letter we propose that the clinically observed diversity in AF progression can be caused by a single process, the progressive stochastic accumulation of transversal cellular uncoupling. Using a simple computer model we show that different time-courses of AF can occur between patients despite a similar degree of transversal cellular uncoupling. Thus the model provides an explanation to the aforementioned clinical observations, namely that (1) the time-course of AF progression can vary significantly between patients, (2) persistent AF can remit to paroxysmal AF, and (3) macrostructurally similar myocardium can show very different AF behaviour as a result of these microstructural differences. In addition to this, the model identifies specific critical architectural patterns of uncoupling between myocytes as the primary cause of AF induction. When access to the microstructure becomes available in the future this insight has the potential to result in patient-specific therapy. 

We have previously developed a model in which the activation wavefronts propagate on an anisotropic structure mimicking the branching network of heart muscle cells \cite{Manani2015} (see supplementary information for a complete description of the model). The tissue is represented by a $L \times L$ square grid of discrete cells where each cell is always coupled to its longitudinal neighbors but with probability $\nu$ to its transversal neighbors. This generates a lattice with anisotropic coupling, mimicking the uncoupling of transversal cell-to-cell connectivity through the parameter $\nu$. We use the simplest model of cell kinetics to mimic the action potential so that a cell may be in one of three states: resting (repolarized), excited (depolarizing), or refractory. An excited cell causes neighboring coupled resting cells to become excited. Thus the wavefront is a coherent propagation of this excitation through the simulated tissue.

For each “patient” the initial conditions are created by assigning the same number of vertical connections (identical initial $\nu$) but at different random positions. Next the accumulation of transversal cellular uncoupling is implemented by reducing $\nu$ (e.g., to mimic the progression of fibrosis or gap junctional uncoupling).  To do this, we run simulations for a period of $T = 4.3 \times 10^7$ time steps in the computer model and vertical connections are removed at a rate of one connection every nine thousand time steps. We note that the actual rate at which transversal uncoupling accumulates in humans is unknown and may differ between patients. Hence, the rates used in the model are set to be identical between simulated “patients” with the aim of capturing the generic phenomenon of the accumulation of uncoupling over time thought to occur frequently in humans \cite{Kottkamp2013}.  We observe the dynamics of activation wavefronts as the transversal uncoupling accumulates in the tissue. All other model parameters are set to physiological values as described in \cite{Manani2015} (see supplementary information). 

We ran 32 lattice simulations, representing 32 patients, with the same initial fraction of vertical connections distributed randomly in each simulated tissue. We start at $\nu = 0.25$ where all the simulated heart muscle tissue are in sinus rhythm (SR) but when lowering $ \nu $, fibrillation may emerge. The number of excited cells can be used to determine when the system is in fibrillation, see Fig. 1 for example and associated electrogram, and hence determine the associated AF burden which we define as the amount of time in AF divided by the total observed time. To define paroxysmal and persistent AF in the model we use a scheme similar to that of Sugihara \textit{et al.} based on AF burden. That is, we call periods of AF burden being 1-99\% paroxysmal AF and   burden of 100\% persistent AF. Permanent AF traditionally refers to the clinical decision not to treat and thus is not informative of the dynamics of AF.

\vspace*{-0.25cm}
\begin{figure}[!ht]\centering
	\includegraphics[width=0.48\textwidth]{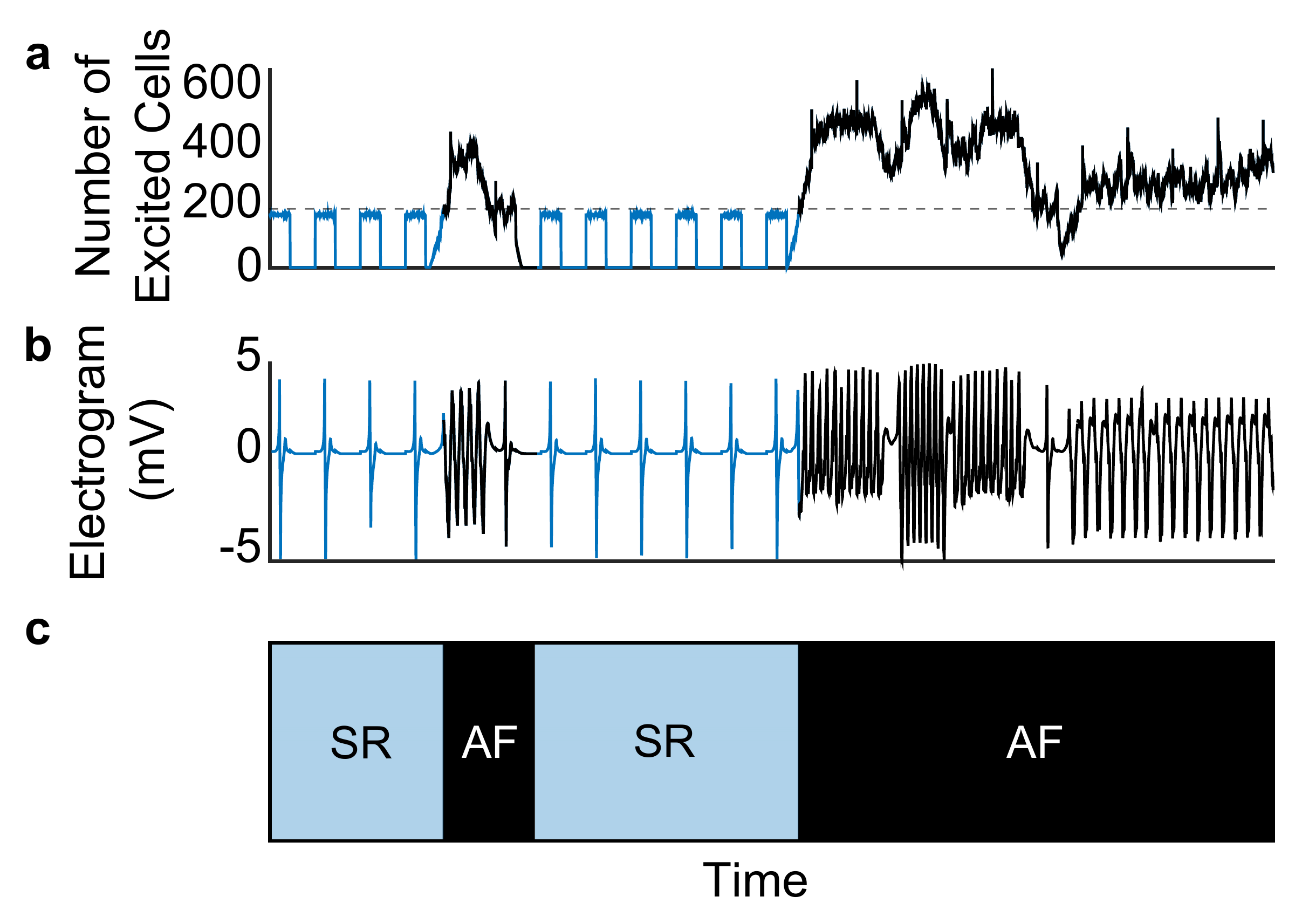}
	\vspace*{-0.5cm}
	\caption{Numerical simulation for a 200 by 200 system where pacemaker cells activate periodically. \textbf{a},  during sinus rhythm (SR, blue curve) the number of excited cells varies with the same period as the pacemaker cells.  When the number of excited cells exceeds a threshold (220, see dashed gray line) it implies that the system is in fibrillation (AF, black curve). The system is defined to return to SR when the system is below threshold for more than one normal sinus rhythm beat.\textbf{ b}, a rectangular electrode of size $1 {mm}^2$  (10 x 10 cells) placed at the center of the tissue is used to simulate the electrogram. During fibrillation (black curve) the rate of the electrogram increases by a factor of 2-5. \textbf{c}, the associated binary signal of the time series in \textbf{a} into periods of SR (blue filled area) and AF (black filled area). Over long-time simulations, the AF burden can be computed from this as the fraction of time in AF.}
	\label{}
\end{figure}

Figure 2 shows the time-course of four particular simulated “patients”. “Patient” A undergoes what would be considered the standard progression from paroxysmal AF, with low AF burden, to persistent AF, with maximal AF burden. “Patient” B, however, shows isolated short-lived episodes of paroxysmal AF with a sudden cross-over to persistent AF. Hence, “Patient” B lacks a gradual progression from paroxysmal AF to persistent AF. “Patient” C also had a few isolated episodes of paroxysmal AF before entering a much more disordered relapsing-remitting phase between paroxysmal and persistent AF with different AF burdens compared to “Patient” B. “Patient” D underwent a sudden transition from sinus rhythm to persistent AF, but has phases of sinus rhythm interrupting persistent AF during the time-course of the disease.

\begin{figure}[!ht]\centering
	\includegraphics[width=0.48\textwidth]{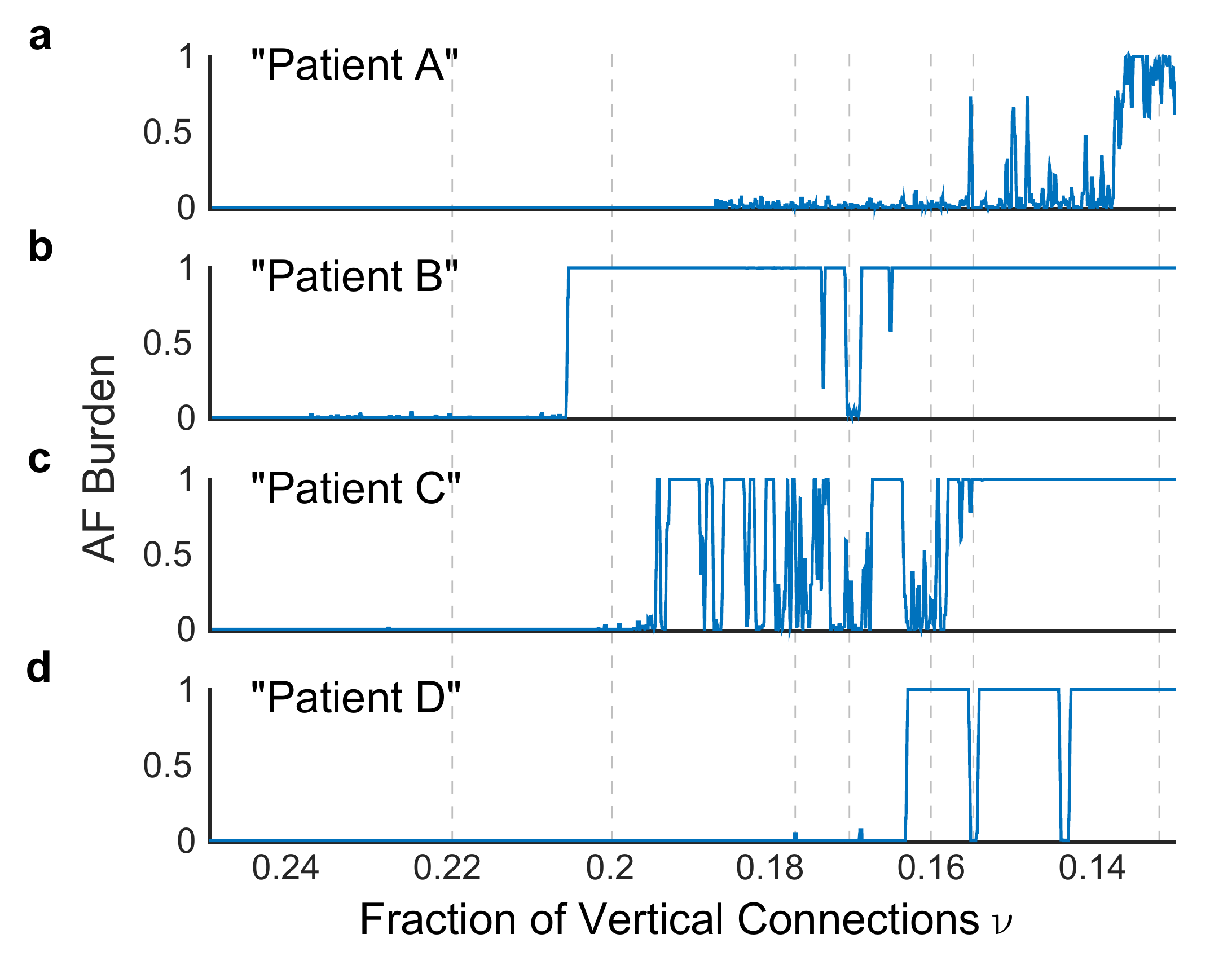}
	\vspace*{-0.5cm}
	\caption{AF burden varies with time, calculated  from the time in AF in a sliding window of $5 \times 10^6$ time steps for four different``patients''. Each simulation begins with an initial fraction of vertical connections of $\nu = 0.25$ and is depleted to $\nu=0.13$ mimicking progression of stochastic uncoupling  over a simulated time period of $ 4.3 \times 10^7$ time steps. \textbf{a},  ``Patient'' A develops paroxysmal AF for $\nu \lesssim 0.188$ which eventually develops into persistent AF at $\nu \lesssim 0.138$. \textbf{b},  ``Patient'' B develops short lived episodes of AF for $0.206 \lesssim \nu \lesssim 0.238$ and a sudden transitions into persistent AF $\nu \approx 0.206$. This relapses into paroxysmal AF three times before remitting back into persistent AF at $\nu \approx 0.166$. \textbf{c},  ``Patient'' C shows a phase of relapsing-remitting paroxysmal to persistent AF for a considerable period of time until AF becomes persistent at $\nu \approx 0.159$. \textbf{d},  ``Patient'' D shows isolated short lived episodes of AF and a sudden transition into persistent AF $\nu \approx 0.164$. Remission back into sinus rhythm occurs twice ($\nu \approx 0.156$ and $\nu \approx 0.145$) before AF becomes persistent. The vertical dashed lines are values of fractions of vertical connections at which simulations are re-run without progressive uncoupling, that is, with a fixed fraction of vertical connections (see Fig. 3).}
	\label{Fig:ModelDynamics}
\end{figure}

These four ``patients'' are archetypes of the time-course of AF that we observe in our simulations. In addition to the variability in the progression to persistent AF we note that the onset of AF occurs at significantly different amounts of uncoupling, that is, fractions of vertical connections. These findings are consistent with the clinical observations that macroscopically similar myocardium can show large variability in AF burden \cite{Sugihara2015,Kottkamp2013}.

Furthermore, we observe that AF activity tends to change suddenly. That is, the frequency and duration of AF episodes change rapidly rather than gradually with progressive uncoupling (see Fig. 3). This is consistent with the clinical observation that macrostructurally similar myocardium (e.g., quantified by the global average fibrosis burden) show very different AF characteristics \cite{Kottkamp2013}. The frequency and duration of AF events are different in each simulated patients. However, in addition to clinical studies we can identify the microscopic origins of the observed behaviour.

\begin{figure}[!ht]\centering
	\includegraphics[width=0.45\textwidth]{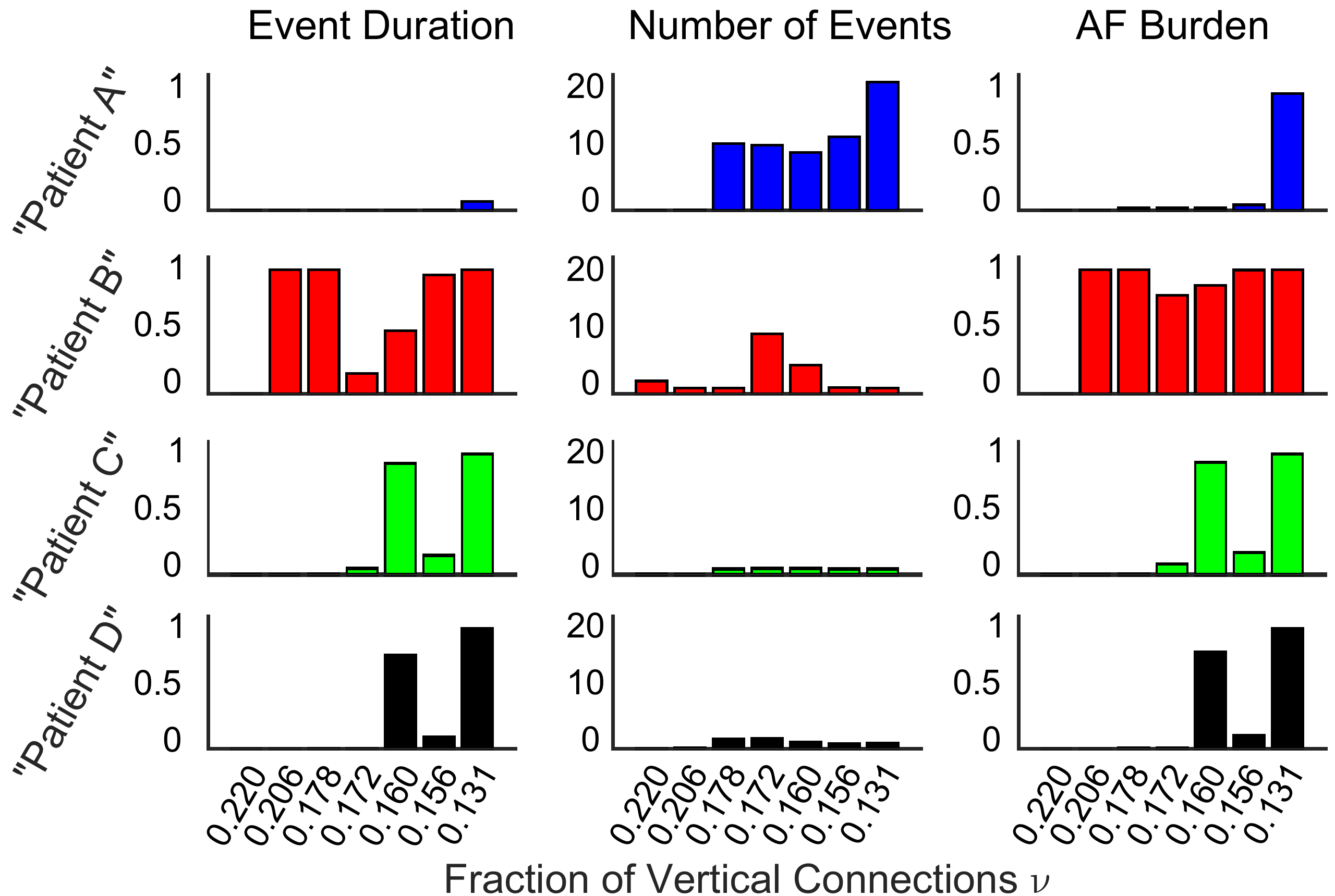}
	\vspace*{-0.1cm}
	\caption{Seven snapshots at $\nu=0.220,$ $0.206,$ $0.178,$ $0.172,$ $0.160,$ $0.156$ and $0.131$ of variability in event duration (expressed as a fraction of the simulation time $1.2 \times 10^7$ time steps), number of AF events and AF burden for repeated simulations as a function of fraction of vertical connections. We re-run the simulations shown in Fig. 2 for $1.2 \times 10^7$ time steps starting from particular values of the fraction vertical connections (see vertical dashed lines in Fig. 2) in which at least one of the four simulations displayed non sinus rhythm behavior. Note that we see variability both within a ``patient'' as the time-course of AF progresses as well as between ``patients''. These three observables of event duration, number of AF events and AF burden are seen to vary significantly between real patients as well \cite{Sugihara2015}. }
	\label{Fig:StoryBoard}
\end{figure}

The differences in the behaviour of these simulations are explained to a large degree by the number of localised critical regions with specific architectural patterns of coupling observed in the simulated tissues. In the model we detect these critical regions by detecting complete loops of wavefront activation, that is, micro-re-entrant circuits. However, these critical patterns of uncoupling might also be determined structurally as these local patches of tissue have connections and dysfunctional cells arranged in a configuration which allows the formation of pinned micro-re-entrant circuits \cite{Manani2015}. These regions are characterised by large contiguous regions of uncoupled cell akin to the obstructive fibrosis found to promote AF in goats \cite{Angel2015}. We observe that the first occurrence of AF coincide with the (chance) emergence of the first critical structure at $\nu = 0.188,0.238,0.228$ and $0.177$ for “Patient” A, B, C and D, respectively, see Fig. 4. Furthermore, the number of these critical regions vary between ``patients'' despite having the same fraction of vertical connections (macroscopic measure). It is the variation in the number of these critical regions that causes the variability in the observed AF behaviour shown in Figures 2 and 3, see Fig. 4 \textbf{e} (see also full set of 32 ``patients'' in supplementary materials).

\begin{figure}[!ht]\centering
	\includegraphics[width=0.48\textwidth]{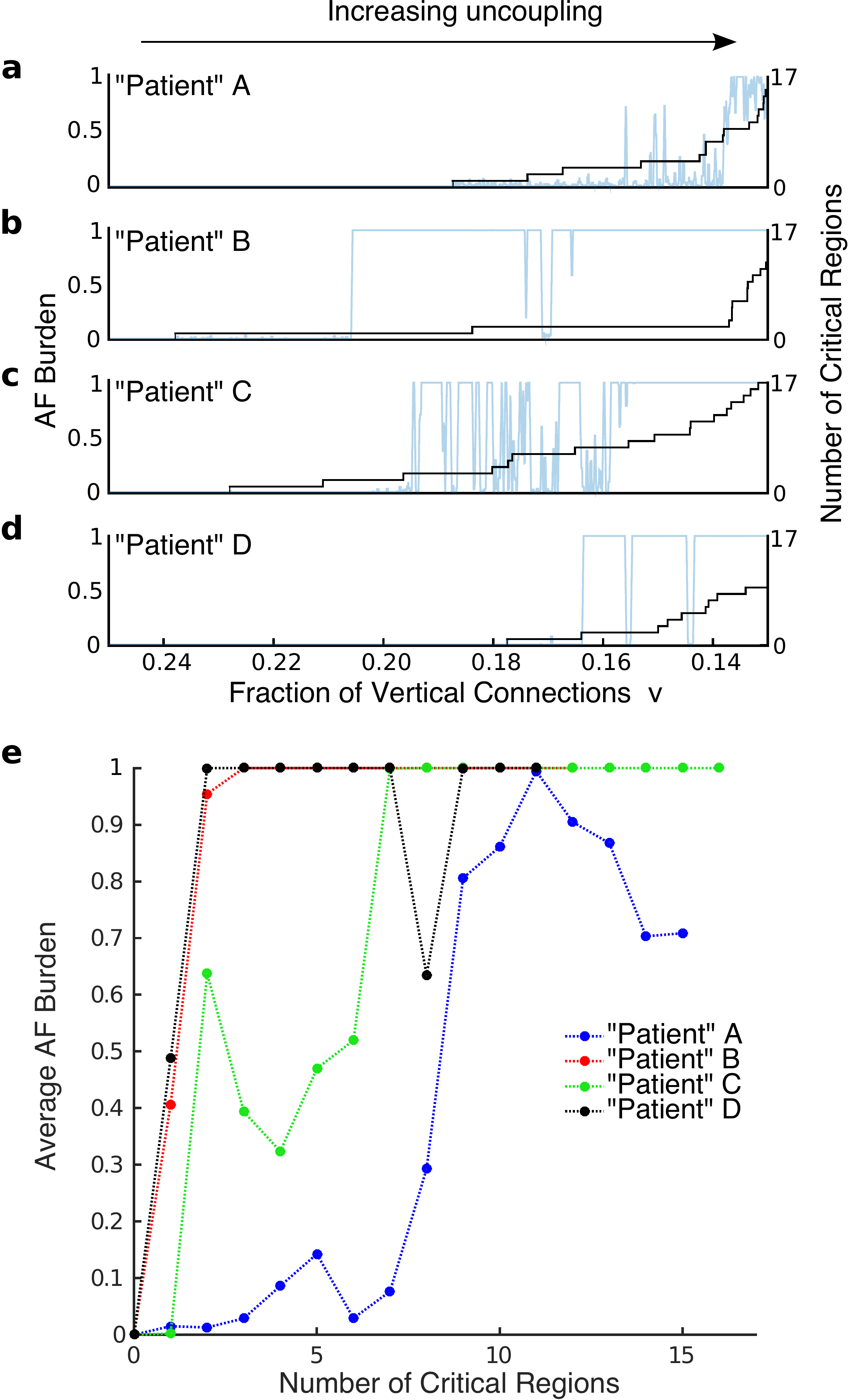}
	\vspace*{-0.55cm}
	\caption{\textbf{a-d}, Accumulation of initiating critical regions (black curve) versus the fraction of vertical connections for ``patients'' A-D. The onset of AF in each ``patient'' coincide with the first appearance of a critical region. As uncoupling progresses (the fraction of vertical connections is reduced), the number of initiating critical regions increases. \textbf{e}, The average AF burden for each ``patient'' as a function of the number of critical regions. The differences in the accumulation of critical regions between ``patients'' better predicts the variability in the AF behavior observed.}
	\label{Fig:TimeInAF}
\end{figure}

The progression of AF and its variability is poorly understood. Fibrosis, among other factors, is known to be important. However, the reason why so much heterogeneity in AF behaviour occurs in patients with similar fibrosis burden is not known. In addition to this, insights from studies in cardiac tissue slices (short time-scales) have yet to be reconciled with the clinical patterns of AF development (long time scales). We bridge this gap using computational modelling and identify how structural characteristics of myocardium and uncoupling alone can give rise to patient variability.
We note that the variability in the simulated patients is strictly due to specific architectural patterns of vertical uncoupling between cells. Fibrosis is one mechanism of cellular uncoupling. An additional mechanism of uncoupling is gap junctional remodelling, whereby the passive high resistance pathways between cells are redistributed to enhance anisotropic conduction and may result in the failure of action potential propagation \cite{Spach2001,Spach2000,Hubbard2007}. Indeed, gap junctional remodelling is known to be arrhythmogenic.

We show that a simple model of heart muscle tissue can display the clinically observed variability in AF progression. This variability originates from the chance occurrence of critical regions characterised by poor vertical connectivity. The model reproduces clinical observations: variability in AF episode onset time, frequency, duration and progression (Fig. 3) along with (1) significant variability in the time-course of AF progression between patients (2) persistent AF remitting to paroxysmal AF and (3) macrostructurally similar myocardium can show very different AF behaviour (Figs. 2 and 4) Thus we show that a single pathological mechanism, namely uncoupling, can result in patterns of AF observed clinically. Specific architectural patterns of uncoupling rather than the global uncoupling (e.g., total fibrosis burden) were observed to drive AF. Thus, our work suggests that the tissue microstructure is essential in determining the time-course of AF in a given patient. This is a first step in relating structural features of myocardium, greatly studied in a basic science context, to patterns of AF in patients, studied in a clinical context. When experimental access to in vivo tissue microstructure becomes available in the future insight from this work might potentially lead to patient-specific therapy.

\begin{acknowledgments}
\textit{Acknowledgements}: We are very grateful to Nick Linton for his pertinent comments regarding the manuscript. This work was supported by the British Heart Foundation (RG/10/11/28457), the ElectroCardioMaths Programme of Imperial BHF Centre of Research Excellence, and the National Institute for Health Research Biomedical Research Centre. K.A.M and K.C conceived the experiment and its interpretation. K.A.M performed the simulations and created the figures. N.S.P provided essential comments in relating the model to a clinical context. K.C. and N.S.P supervised the project. All authors discussed the results, commented on and contributed to the writing of the manuscript. 
\end{acknowledgments}

\bibliographystyle{apsrev}
\bibliography{References}

\end{document}